\documentclass[fleqn,usenatbib]{mnras}

\pdfoutput=1
\usepackage{graphicx,subfigure,bm,color,psfrag,hyperref}
\usepackage{amsfonts}
\usepackage{lipsum}
\usepackage{mathtools}
\usepackage{verbatim}
\usepackage{color}
\usepackage{newtxtext,newtxmath}
\usepackage[T1]{fontenc}
\usepackage{amsmath} 

\DeclareRobustCommand{\VAN}[3]{#2}
\let\VANthebibliography\thebibliography
\def\thebibliography{\DeclareRobustCommand{\VAN}[3]{##3}\VANthebibliography}

\begin{document}

%-----------------------------------------------------

\title[Interacting Dark Energy in a closed universe]{Interacting Dark Energy in a closed universe}

%----------------------------------------------------

\author[Eleonora Di Valentino, Alessandro Melchiorri, Olga Mena, Supriya Pan and Weiqiang Yang]{
Eleonora Di Valentino $^{1}$\thanks{E-mail: eleonora.divalentino@manchester.ac.uk}
Alessandro Melchiorri $^{2}$
Olga Mena $^{3}$ 
Supriya Pan $^{4}$
and Weiqiang Yang $^{5}$\\
$^{1}$Jodrell Bank Center for Astrophysics, School of Physics and Astronomy, 
University of Manchester, Oxford Road, Manchester, M13 9PL, UK\\
$^{2}$Physics Department and INFN, Universit\`a di Roma ``La Sapienza'', Ple Aldo Moro 2, 00185, Rome, Italy\\
$^{3}$IFIC, Universidad de Valencia-CSIC, 46071, Valencia, Spain\\
$^{4}$Department of Mathematics, Presidency University, 86/1 College Street, Kolkata 700073, India\\
$^{5}$Department of Physics, Liaoning Normal University, Dalian, 116029, People's Republic of China
}

\date{Accepted XXX. Received YYY; in original form ZZZ}

\pubyear{2020}

\label{firstpage}
\pagerange{\pageref{firstpage}--\pageref{lastpage}}
\maketitle

%------------------------------------------------------------

\begin{abstract}
Recent measurements of the Cosmic Microwave Anisotropies power spectra measured by the Planck satellite show a preference for a closed universe at more than $99 \%$ Confidence Level (CL). Such a scenario is however in disagreement with several low redshift observables, including luminosity distances of Type Ia Supernovae. Here we show that Interacting Dark Energy (IDE) models can ease the discrepancies between Planck and Supernovae Ia data in a closed Universe, leading to a preference for both a coupling and a curvature different from zero above the 99\% CL. Therefore IDE cosmologies remain as very appealing scenarios, as they can provide the solution to a number of observational tensions in different fiducial cosmologies. The results presented here strongly favour broader analyses of cosmological data, and suggest that relaxing the usual flatness and vacuum energy assumptions can lead to a much better agreement among theory and observations.  
\end{abstract}

%\pacs{98.80.-k, 95.36.+x, 95.35.+d, 98.80.Es}
%--------------------------------------------------------
\begin{keywords}
cosmic background radiation -- cosmological parameters -- dark energy -- observations
\end{keywords}
%--------------------------------------------------------

\section{Introduction}

The recent Planck Cosmic Microwave Background (CMB) power spectra, when analyzed with the official \emph{Plik} likelihood, show a clear preference for a closed Universe at more than three standard deviations~\cite{Aghanim:2018eyx,Handley:2019tkm,DiValentino:2019qzk,DiValentino:2020hov}. This result clearly introduces a problem for the standard $\Lambda$CDM cosmological scenario, based on the inflationary prediction of a flat universe. Undetected systematics can play a role and it is obviously too soon to exclude a flat universe. For example, the authors of the \emph{CamSpec} alternative likelihood claim a larger value of the $\chi^2$ fit for closed universes. However, the marginalized constraints obtained from this alternative likelihood still prefer closed models at a significant level (larger than $99 \%$ CL). 

While the compatibility with a closed universe of the Planck dataset is solid, a major problem of models with positive curvature is that they further exacerbate to at least $5.4\sigma$~\cite{DiValentino:2019qzk}, the already $4.4\sigma$ tension on the Hubble constant between the $H_0$ value measured by Planck~\cite{Aghanim:2018eyx} in a $\Lambda$CDM model and the value obtained by the SH0ES collaboration~\cite{Riess:2019cxk} (R19).~\footnote{For a recent overview of the $H_0$ tension, see Ref.~\cite{DiValentino:2020zio}.} 
Furthermore, introducing a curvature in the universe free to vary implies a tension between Planck CMB and Baryon Acoustic Oscillation (BAO) data~\cite{DiValentino:2019qzk,Handley:2019tkm} and also between Planck CMB observations and the full-shape galaxy power spectrum~\cite{Vagnozzi:2020zrh}.

The authors of Ref.~\cite{DiValentino:2020hov} presented the possibility of solving the tension between Planck and Supernovae Ia luminosity distance measurements within a closed universe by including phantom dark energy, ruling out both the flatness and the cosmological constant scenario at more than 99\% CL, while the tension with the BAO data was still persistent. 

Clearly, the tension with BAO is a major problem for these closed-phantom models. However one could argue that the BAO reconstruction from galaxy data (that is performed under the assumption of $\Lambda$CDM) could be affected by a radically different choice of the dark energy component.
It is therefore interesting to consider more physically motivated dark energy models that could induce an effective phantom behaviour in the context of a closed universe.

In this manuscript we 
consider a model of Dark Energy (DE) interacting with the Dark Matter (DM) as a possibility for solving the tension between Planck and the luminosity distance measurements within a non-flat universe. The IDE scenarios \cite{Bolotin:2013jpa,Wang:2016lxa,Pettorino:2013oxa,Yang:2017zjs}) became very promising due to an evidence of a non-null interaction \cite{Salvatelli:2014zta,Kumar:2017dnp,DiValentino:2017iww,Kumar:2019wfs}  and 
recently got plenty of attention as alternative scenarios for solving the Hubble constant tension, see for example
Refs.~\cite{DiValentino:2019ffd,DiValentino:2019jae,Pan:2020bur,Gomez-Valent:2020mqn,Lucca:2020zjb,Yang:2020uga,Yang:2019uog,Yang:2018ubt} for updated results with Planck 2018, and references therein with older data. This happens because the flux of energy from the DM sector to the DE one within some IDE scenarios naturally provides a higher value of the Hubble constant $H_0$ to compensate for both the lowering of the matter energy density and the shift of the acoustic peaks in the damping tail (see also~\cite{DiValentino:2020leo} for a discussion about degeneracies in the parameters). 
Therefore, it is timely to explore whether IDE scenarios can reconcile the discrepancies between Planck and Supernovae Ia luminosity distance measurements within non-flat cosmologies. 

The paper is organized as follows. In Sec.~\ref{model} we present the IDE model analyzed here. Section~\ref{method} presents the methodology and the datasets used in this work, while in Sec.~\ref{results} we show the results obtained in our analysis and discuss their physical implications. Finally, Sec.~\ref{conclu} summarizes the conclusions of the present work.

\begin{table}
\caption{Flat priors on the main cosmological parameters used in this work. }
\begin{center}
\begin{tabular}{|c|c|c|}
\hline
Parameter                    & prior \\
\hline
$\Omega_{\rm b} h^2$         & $[0.005,0.1]$ \\
$\Omega_{\rm c} h^2$         & $[0.001,0.99]$ \\
$100\theta_{MC}$             & $[0.5,10]$ \\
$\tau$                       & $[0.01,0.8]$ \\
$n_\mathrm{S}$               & $[0.7,1.3]$ \\
$\log[10^{10}A_{s}]$         & $[1.7, 5.0]$ \\
$\Omega_k$                   & $[-3,3]$ \\
$\xi$                        & $[-1,0]$ \\
\hline %S
\end{tabular}
\end{center}
\label{priors}
\end{table}

\section{Interacting dark energy overview}
\label{model}

In the context of a Friedmann-Lema\^{i}tre-Robertson-Walker (FLRW) universe allowing spatial curvature, we consider a very generalized cosmic scenario where the two main dark fluids of the universe, namely a pressureless DM and a DE component share an energy and/or momentum exchange mechanism in a non-gravitational way. The energy densities of the pressureless DM and DE are respectively denoted by $\rho_c$ and $\rho_x$ and additionally we assume that DE has a constant equation-of-state $w$. At the background level, the conservation equations for the pressureless DM and DE components can be decoupled into two separate equations with an inclusion of an arbitrary function, $Q$, known as the coupling or interacting function, as follows: 

\begin{eqnarray}
\label{cons-CDM}
\dot{\rho}_c+3{\cal H}\rho_c &=& Q\,, \\
\label{cons-DE}
\dot{\rho}_x+3{\cal H}(1+w)\rho_x &=&-Q\,,
\end{eqnarray}
where the dot denotes the differentiation with respect to the conformal time $\tau$, and ${\cal H} \equiv \dot{a}/a$ refers to the conformal Hubble rate of the FLRW universe. The function $Q$ determines the direction of energy (and/or momentum)
transfer between the dark sectors through its sign. For instance, $Q>0$ ($Q < 0$) indicates the transfer of energy (and/or momentum) from DE (DM) to DM  (DE). Once a specific function for the interaction rate $Q$ is prescribed, by solving either analytically or numerically the continuity equations Eq.~(\ref{cons-CDM}) and Eq.~(\ref{cons-DE}) the evolution of the universe will be fully determined, as all the other fluids are obeying the usual conservation equations, that is, they do not take part in the interaction process. Therefore, the interaction function plays a crucial role in the determination of the cosmic dynamics. Usually, there is no final form of the interaction function in the literature, but there are some well-known interaction functions. This is the case of the IDE model described by

\begin{eqnarray}
Q = \xi{\cal H}\rho_x\,,
\label{eq:coupling}
\end{eqnarray}
where $\xi$ is the dimensionless coupling parameter
which characterizes the strength of the interaction between the dark sectors. The coupling function in Eq.~(\ref{eq:coupling}) was initially proposed purely from a phenomenological perspective. While most of the interaction functions are phenomenological, some recent investigations have shown that the model in Eq.~(\ref{eq:coupling}) can be derived from some multi-scalar field action~\cite{Pan:2020zza}. Additionally, apart from the scalar field theory, the origin of various interaction functions can also be motivated from other existing cosmological theories with a Lagrangian description~\cite{Pan:2020mst}. Finally, we note that the model can be analytically solved leading to the closed form expressions for $\rho_c$ and $\rho_x$ \cite{DiValentino:2019jae}. The coincidence parameter $r = \rho_c/\rho_x$ also assumes an analytic expression and for $z \rightarrow 0$, $r$ becomes a constant, thus, alleviating the coincidence problem.   

The interaction term $Q$ also affects the perturbation equations. In the context of linear perturbation theory, assuming the synchronous gauge, one can write down the density perturbation $\delta$ and the velocity divergence $\theta$ of the dark fluids as~\cite{Valiviita:2008iv,Gavela:2009cy,Gavela:2010tm,DiValentino:2019ffd}:

\begin{eqnarray}
\label{eq:deltac}
\dot{\delta}_c &=& -\theta_c - \frac{1}{2}\dot{h} +\xi{\cal H}\frac{\rho_x}{\rho_c}(\delta_x-\delta_c)+\xi\frac{\rho_x}{\rho_c} \left ( \frac{kv_T}
{3}+\frac{\dot{h}}{6} \right )\,, \\
\label{eq:thetac}
\dot{\theta}_c &=& -{\cal H}\theta_c\,,\\
\label{eq:deltax}
\dot{\delta}_x &=& -(1+w) \left ( \theta_x+\frac{\dot{h}}{2} \right )-\xi \left ( \frac{kv_T}{3}+\frac{\dot{h}}{6} \right ) \nonumber \\
&&-3{\cal H}(1-w) \left [ \delta_x+\frac{{\cal H}\theta_x}{k^2} \left (3(1+w)+\xi \right ) \right ]\,,\\
\label{eq:thetax}
\dot{\theta}_x &=& 2{\cal H}\theta_x+\frac{k^2}{1+w}\delta_x+2{\cal H}\frac{\xi}{1+w}\theta_x-\xi{\cal H}\frac{\theta_c}{1+w}\,.
\end{eqnarray}
 The governing equations both at the background and perturbation levels completely determine the dynamics of the interacting universe.

Finally, we make an important comment on the early time instabilities which are associated with the interacting cosmic scenarios. As already noticed, the interaction function introduces a new free parameter $\xi$, which controls the energy flow between the  dark sectors and along with the dark energy equation of state parameter, $w$, it plays a very active part in the modified perturbation equations to determine whether the interaction model leads to early time instabilities or not. This problem has been examined by several works in the past~\cite{Gavela:2009cy,Gavela:2010tm} which led to the conclusion that the instability problem can be avoided if the signs of $\xi$ and $(1+w)$ are different. Therefore, in the present article we have considered the opposite signs for $\xi$ and $(1+w)$ in order to ensure that the underlying model does not suffer from early time instabilities, fixing the dark energy equation of state parameter $w =-0.999$.

\begin{figure*}
\begin{center}
	\includegraphics[width=0.55\linewidth]{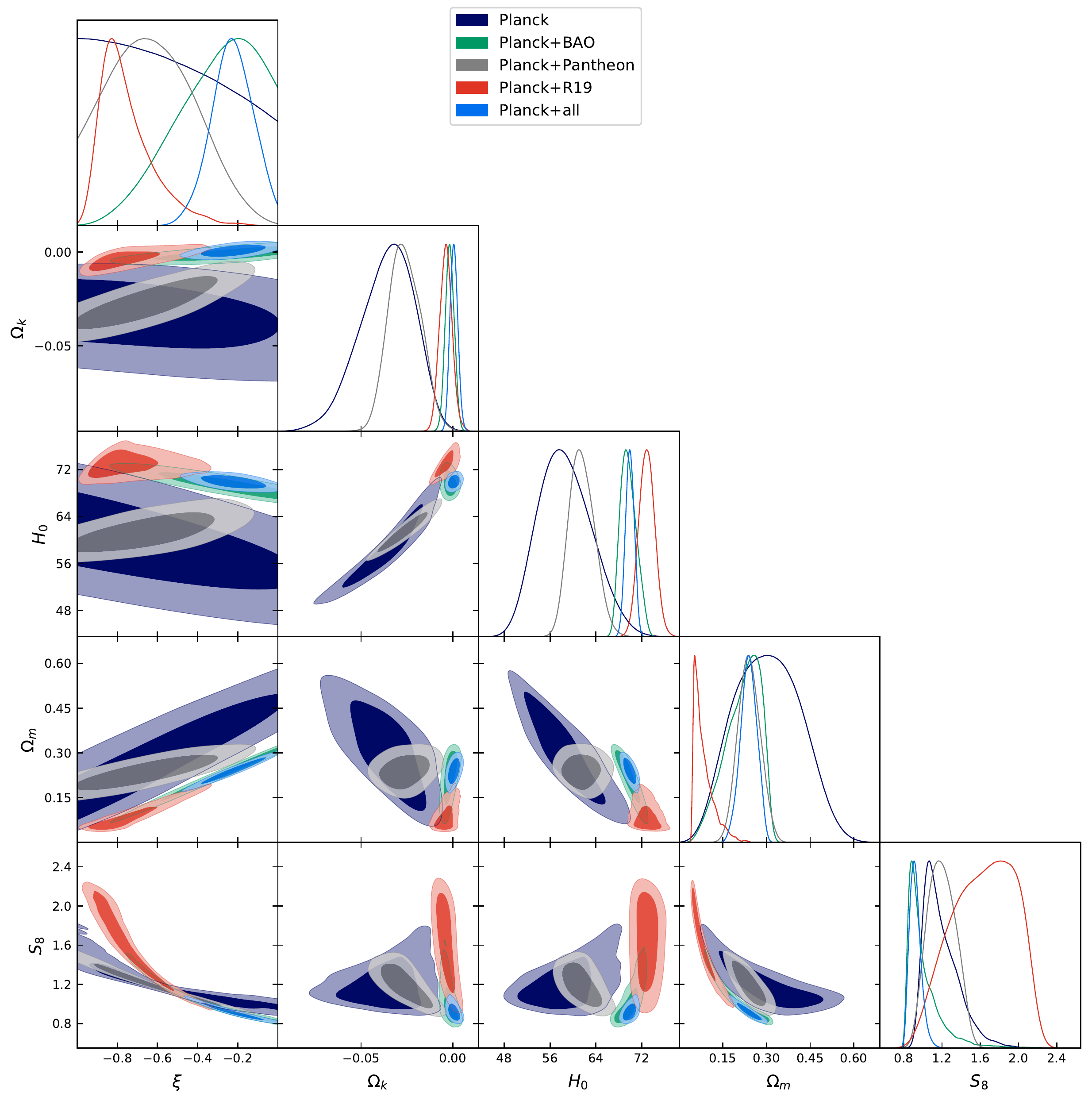}
	\caption{One dimensional posterior distributions and two-dimensional joint contours at 68\% and 95\% CL for the IDE+$\Omega_k$ model.}
	\label{fig:help}
\end{center}
\end{figure*}

\begin{center}                  
\begin{table*}
\caption{Observational constraints at 68$\%$~CL on the independent %(above the line) 
and derived %(below the line)
cosmological parameters arising from analyses to Planck observations within the $\Lambda$CDM, $\Lambda$CDM+$\Omega_k$ and IDE+$\Omega_k$ cosmologies. In the bottom line we quote the $\ln B_{ij}$ computed with respect to the $\Lambda$CDM cosmology. The positive values are indicating a preference for the models different from the $\Lambda$CDM one.}
\scalebox{0.8}{
\begin{tabular}{cccccccccccccccc}  
\hline\hline                        
Parameters & $\Lambda$CDM   & IDE & $\Lambda$CDM+$\Omega_k$ & IDE+$\Omega_k$ \\ \hline
 
 $\Omega_{\rm b} h^2$ & $    0.02236 \pm 0.00015$ &  $    0.02239\pm0.00015$ & $    0.02260\pm0.00017$ & $    0.02261\pm 0.00017$  \\
 
$\Omega_{\rm c} h^2$ & $    0.1202\pm0.0014$  & $    <0.0634$ & $    0.1181\pm0.0015$ & $    0.077^{+0.035}_{-0.019} $\\

$100\theta_{\rm MC}$ & $    1.04090\pm 0.00031$ &  $    1.0458^{+0.0033}_{-0.0021}$ &  $    1.04116\pm 0.00033$ & $    1.0437^{+0.0012}_{-0.0023}$ \\

$\tau$ & $    0.0544^{+0.0070}_{-0.0081}$ &  $    0.0541\pm0.0076$ &  $    0.0486\pm 0.0082$ & $    0.0481^{+0.0085}_{-0.0076}$ \\

$n_s$ & $0.9649\pm0.0044$ & $0.9655\pm0.0043$ & $0.9706\pm0.0048$ & $0.9708\pm0.0047$  \\

${\rm{ln}}(10^{10}A_s)$ & $3.045\pm0.016$ & $3.044\pm0.016$ & $3.028\pm0.017$ & $3.027^{+0.017}_{-0.016}$ \\

$\xi$ & [0]  & $-0.54^{+0.12}_{-0.28}$ & [0] & $<-0.385$  \\

$\Omega_k$ & [0] & [0] & $-0.044^{+0.018}_{-0.015}$ & $-0.036^{+0.017}_{-0.013}$  \\

\hline

$H_0 $[(km/s)/Mpc] & $   67.27\pm0.60$&  $   72.8^{+3.0}_{-1.5}$ & $   54.4^{+3.3}_{-4.0}$ & $   58.7^{+4.1}_{-5.2}$ \\

$\sigma_8$ & $    0.8120\pm0.0073$ &  $    2.3^{+0.4}_{-1.4}$ & $    0.744\pm0.015$ & $    1.31^{+0.10}_{-0.54}$ \\

$\Omega_m$ & $0.3166\pm0.0084$ & $0.139^{+0.034}_{-0.095}$ & $0.485^{+0.058}_{-0.068}$ & $0.30\pm0.11$  \\

$S_8$ & $    0.834\pm0.016$ &  $    1.30^{+0.17}_{-0.44}$ & $  0.981\pm0.049  $ & $    1.20^{+0.10}_{-0.22}$\\
\hline
%$\chi_{\rm bestfit}^2$ & $    2772.65 $ &  $ 2770.88  $ & $ 2760.09   $& $ 2761.94  $\\
$\ln B_{ij}$& $    - $ &  $ 1.2  $ & $ 2.3   $& $ 2.5  $\\
\hline\hline                         
\end{tabular} }
\label{tab:comparison}                      
\end{table*}                         
\end{center}

\begin{center}   
\begin{table*}
\caption{Observational constraints at 68$\%$~CL on the independent 
%(above the line) 
and derived 
%(below the line) 
cosmological parameters arising from analyses to different data combinations within the IDE+$\Omega_k$ model. The $\ln B_{ij}$ are computed with respect to the $\Lambda$CDM+$\Omega_k$ cosmology for the very same dataset combination. The negative values are indicating a preference for the $\Lambda$CDM+$\Omega_k$ scenario, while the positive values for the IDE+$\Omega_k$ model. }
\scalebox{0.8}{
\begin{tabular}{cccccccccccccccc}  
\hline\hline                        
Parameters & Planck   & Planck & Planck& Planck &  Planck \\ 
 & & +BAO  & + Pantheon & + R19 & + all\\ \hline
 
 $\Omega_{\rm b} h^2$ &$    0.02261\pm 0.00017$& $    0.02241 \pm 0.00016$ &  $    0.02258\pm0.00016$ & $    0.02247\pm0.00016$ & $    0.02239\pm 0.00015$  \\
 
$\Omega_{\rm c} h^2$ & $    0.077^{+0.035}_{-0.019} $&$    0.082^{+0.033}_{-0.015}$  & $   0.068^{+0.013}_{-0.018}$ & $    <0.0253$ & $    0.093^{+0.013}_{-0.011} $\\

$100\theta_{\rm MC}$ &$    1.0437^{+0.0012}_{-0.0023}$& $    1.04327^{+0.00009}_{-0.00022}$ &  $    1.0442^{+0.0012}_{-0.0010}$ &  $    1.0480^{+0.0020}_{-0.0008}$ & $    1.04249^{+0.00074}_{-0.00086}$ \\

$\tau$ & $    0.0481^{+0.0085}_{-0.0076}$&$    0.0541\pm0.0081$ &  $    0.0495\pm0.0080$ &  $    0.0534\pm 0.0079$ & $    0.0542\pm0.0079$ \\

$n_s$ & $0.9708\pm0.0047$&$0.9662\pm0.0047$ & $0.9701\pm0.0046$ & $0.9679\pm0.0046$ & $0.9653\pm0.0047$  \\

${\rm{ln}}(10^{10}A_s)$ & $3.027^{+0.017}_{-0.016}$& $3.043\pm0.016$ & $3.031\pm0.017$ & $3.040\pm0.016$ & $3.045\pm0.016$ \\

$\xi$& $<-0.385$ & $-0.32^{+0.31}_{-0.09}$  & $-0.62^{+0.19}_{-0.25}$ & $-0.75^{+0.06}_{-0.16}$ & $-0.23\pm0.10$  \\

$\Omega_k$& $-0.036^{+0.017}_{-0.013}$ & $-0.0016\pm0.0024$ & $-0.0261\pm0.0087$ & $-0.0038\pm0.0034$ & $0.0006\pm0.0021$  \\

\hline

$H_0 $[(km/s)/Mpc] & $   58.7^{+4.1}_{-5.2}$& $   69.7^{+1.2}_{-1.6}$&  $   61.6^{+2.0}_{-2.4}$ & $   72.9\pm1.4$ & $   69.93\pm0.75$ \\

$\sigma_8$ & $    1.31^{+0.10}_{-0.54}$& $    1.27^{+0.04}_{-0.46}$ &  $    1.36^{+0.20}_{-0.31}$ & $    3.4^{+1.2}_{-1.4}$ & $    1.04^{+0.08}_{-0.15}$ \\

$\Omega_m$ & $0.30\pm0.11$& $0.219^{+0.076}_{-0.040}$ & $0.240\pm0.038$ & $0.084^{+0.010}_{-0.039}$ & $0.239\pm0.028$  \\

$S_8$& $    1.20^{+0.10}_{-0.22}$ & $    1.01^{+0.04}_{-0.18}$ &  $    1.20^{+0.14}_{-0.16}$ & $  1.64^{+0.41}_{-0.27}  $ & $    0.921^{+0.043}_{-0.069}$\\
\hline
$\ln B_{ij}$& $ 0.2 $ &  $ -1.0  $ & $ 3.2   $& $ 5.8  $ & $ -0.4 $\\
\hline\hline                         
\end{tabular}} 
\label{tab:IDE+omegak}                      
\end{table*}                         
\end{center}

\section{Observational data and Methodology}
\label{method}

We consider a baseline IDE+$\Omega_k$ model described by eight cosmological parameters. These will be the baryon energy density $\Omega_{\rm b}h^2$, the cold dark matter energy density $\Omega_{\rm c}h^2$, the ratio between the sound horizon and the
angular diameter distance at decoupling $\theta_{MC}$, the reionization optical depth 
$\tau$, the amplitude of the scalar primordial power spectrum 
$A_{s}$, the spectral index $n_s$, the dimensionless coupling $\xi$ and the curvature parameter $\Omega_k$. 

To analyse this interacting scenario with curvature and also to derive the cosmological constraints, we make use of the combinations of the most recent observational data from various sources listed below:

\begin{itemize}
    \item {\bf Planck}: we consider as a baseline dataset the latest Cosmic Microwave Background (CMB) measurements provided by the final 2018 Planck legacy release \cite{Aghanim:2018eyx,Aghanim:2019ame}.
    
    \item {\bf BAO}: we include a compilation of Baryon Acoustic Oscillations (BAO) measurements from different experiments, namely 6dFGS~\cite{Beutler:2011hx}, SDSS-MGS~\cite{Ross:2014qpa}, and BOSS DR12~\cite{Alam:2016hwk} surveys, as used by the Planck collaboration in Ref.~\cite{Aghanim:2018eyx}. 
    \item {\bf Pantheon}: we use the 1048 data points in the redshift region $z \in [0.01, 2.3]$ of the luminosity distance data of type Ia Supernovae from the Pantheon catalog~\cite{Scolnic:2017caz}. 
    
    \item {\bf R19}: we add a Gaussian prior on the Hubble constant as estimated from a reanalysis of the Hubble Space Telescope data using Cepheids as calibrators by the SH0ES collaboration in 2019~\cite{Riess:2019cxk}, i.e. $H_0 = 74.03 \pm 1.42$ km/s/Mpc at $68\%$ CL.

\end{itemize}

For the analysis, we use our modified version of the publicly available Markov Chain Monte Carlo code \texttt{CosmoMC}~\cite{Lewis:2002ah,Lewis:1999bs} package (see \url{http://cosmologist.info/cosmomc/}). This code supports the 2018 Planck likelihood~\cite{Aghanim:2019ame}, implements an efficient sampling of the posterior distribution using the fast/slow parameter decorrelations \cite{Lewis:2013hha}, and has a convergence diagnostic based on the Gelman-Rubin statistics~\cite{Gelman:1992zz}.

\begin{figure}
\begin{center}
	\includegraphics[width=0.72\linewidth]{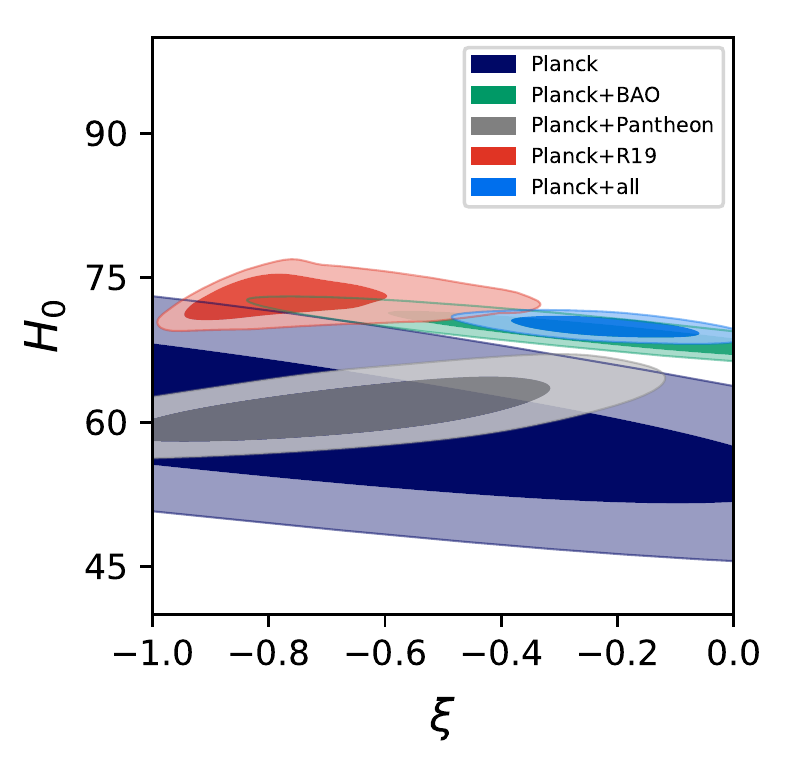}
	\includegraphics[width=0.76\linewidth]{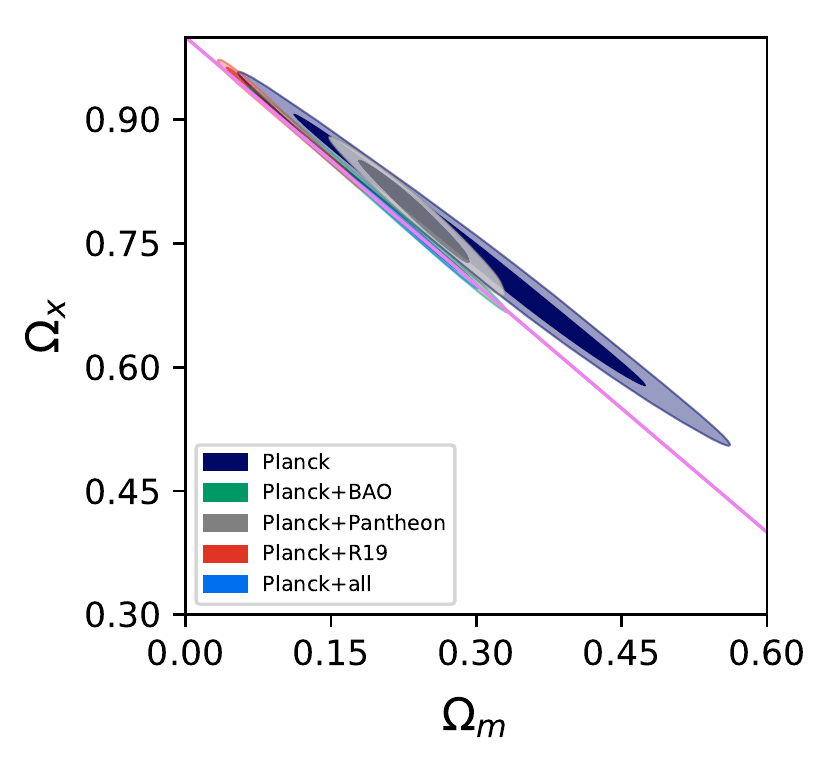}
	\caption{Two-dimensional contour plots at 68\% and 95\% CL for the IDE+$\Omega_k$ model, showing the planes ($H_0$,$\xi$) in the top panel and ($\Omega_\Lambda$,$\Omega_m$) in the bottom one, where the magenta line corresponds to the flatness scenario.}
	\label{fig:help2}
\end{center}
\end{figure}

\section{Results}
\label{results}

We show in Tab.~\ref{tab:comparison} a comparison of the constraints on the cosmological parameters obtained within four different scenarios: the flat $\Lambda$CDM standard model, the flat IDE model, the $\Lambda$CDM plus curvature model, and the IDE plus curvature model. 

The first thing that we can notice is that the combination of the IDE model with a curvature of the universe softens the critical peculiarities we have for the two models separately (see Tab.~\ref{tab:comparison}), as for example the particular lower/higher values for $H_0$/$\Omega_m$ in the $\Lambda$CDM+$\Omega_k$ scenario, or the strong evidence for $\xi$ associated to an exceptional lower amount of dark matter in the IDE model. Indeed, within the IDE+$\Omega_k$ scenario we find for Planck a much more reasonable (larger) value for the matter density $\Omega_m$, see also Fig.~\ref{fig:help}, that corresponds to a bound on $\Omega_c h^2$, instead of just obtaining an upper limit, as in the case of flat IDE cosmologies (see Tab.~\ref{tab:comparison}). 
A word of caution is needed here. While the values of the present matter density $\Omega_m$ may be argued to be very small and incompatible with structure formation processes, we remind the reader that within interacting cosmologies the growth of dark matter perturbations will be larger than in uncoupled models. This feature will be general for models with negative coupling and in which the energy exchange among the dark sectors is proportional to $\rho_x$, due to a suppression of the friction term and an enhancement of the source term in the differential growth equation, see e.g. Refs.~\cite{Honorez:2010rr,CalderaCabral:2009ja}. While this statement holds at the linear perturbation level, a very similar behaviour will be expected at mildly non-linear smaller scales.

Notice that the preference for a closed universe persists in the IDE+$\Omega_k$ case at more than 99\% CL, but with a slightly larger value for the Hubble constant, lowering the tension with R19 at $3.6\sigma$ due to the larger error bars. These two effects can be clearly noticed from the results depicted in Fig.~\ref{fig:help}, which illustrates the one and two-dimensional posterior probability distributions for some of the most interesting parameters. Finally, the improvement in the fit for the IDE+$\Omega_k$ model is significant, as we can see from the logarithm of the Bayes factor, $B_{ij}$ with respect to the $\Lambda$CDM model (computed using the \texttt{MCEvidence} code~\cite{Heavens:2017hkr,Heavens:2017afc}) that is equal to $\ln B_{ij}=2.5$, i.e. a {\it definite} evidence according to the revised Jeffreys scale~\cite{Kass:1995loi}.

We present in Tab.~\ref{tab:IDE+omegak} the bounds on the IDE+$\Omega_k$ model for several combinations of the datasets. The most evident result is that for the IDE+$\Omega_k$ scenario Planck and Pantheon datasets are in excellent agreement, leading to a preference for a curvature component and a coupling different from zero with a significance above 99\% CL (see Fig.~\ref{fig:help2}), and a Bayes factor equal to $\ln B_{ij}=3.2$, indicating a {\it strong} evidence for this scenario with respect to the $\Lambda$CDM +$\Omega_k$ case.
However, this agreement is happening at the price of raising again the $H_0$ tension with R19 to the $5.2\sigma$ level, and it is also in strong disagreement with the BAO data, as can be noticed from the left panel of Fig.~\ref{fig:help2}. Indeed, the Planck+BAO dataset combination prefers a flat universe within 68\% CL and a coupling different from zero at about $1\sigma$, while the Planck+R19 prefers a closed universe at about one standard deviation and a coupling at more than 99\% CL. 
We can, however, note that the constraints from Planck+R19 on the matter density and the amplitude of scalar perturbations, albeit at the linear level, are probably unrealistic even considering the word of caution stressed before. 
The strong improvement in the evidence of Tab~\ref{tab:IDE+omegak}
for this case should, therefore, be considered with some grain of salt since the inclusion of clustering data could strongly play against this solution. In a few words, the simple $\Omega_k$+IDE scenario considered here does not solve the Hubble tension, so we need to consider further extensions.

\section{Conclusions}
\label{conclu}

In this paper we consider an extension of the standard $\Lambda$CDM model by introducing simultaneously a non-gravitational interaction between DE and DM together with a curvature component in our universe. The aim is to investigate whether the very same IDE scenario, strongly motivated for solving the $H_0$ tension, is also able to solve the existing tensions between high and low redshift observations within non-flat cosmologies (namely, the very low value for Hubble constant $H_0$ obtained within the $\Lambda$CDM+$\Omega_k$ scenario). Therefore, this is a very general scheme, which, in one hand, extends the non-interacting cosmological scenarios including the $\Lambda$CDM as the base model, and, on the other hand, includes the very interesting and timely possibility of a curvature component in our universe. Such a possibility has recently been strengthened from the recent observational evidences. Our analyses confirm previous findings within the simpler non-flat $\Lambda$CDM picture: Planck observations prefer a positive curvature of the universe at more than 99\% CL, but this exacerbates the Hubble constant tension at more than $5\sigma$.  While non-flat IDE scenarios provide a larger value of $H_0$, the tension is still present  with a significance of $3.6\sigma$.

Nevertheless  other forms for the interaction function and for the equation of state of the dark energy component could further alleviate  this tension, as it is easened in the right direction due to the much lower value of $\Omega_m$ required within some family of IDE models. Interactions among the dark sectors of our universe therefore remain as a very appealing scenarios, as they can provide the solution to a number of cosmological tensions in different fiducial models.

%----------------------------------------------------
\section*{Acknowledgements}
EDV acknowledges support from the European Research Council in the form of a Consolidator Grant with number 681431. OM is supported by the Spanish grants FPA2017-85985-P, PROMETEO/2019/083 and by the European ITN project HIDDeN (H2020-MSCA-ITN-2019//860881-HIDDeN). SP acknowledges the Science and Engineering Research Board under its MATRICS Scheme (File No. MTR/2018/000940), Govt. of India. WY acknowledges the National Natural Science Foundation of China (Grants No. 11705079 and No. 11647153). 

%--------------------------------------------------------
\section*{Data Availability}
We used the publicly available cosmological probes, such as the CMB power spectra from Planck 2018, BAO data from 6dFGS, SDSS-MGS, BOSS DR12, Hubble constant measurement from the SH0ES collaboration, and the 
Pantheon catalogue. 
%--------------------------------------------------------

\bibliographystyle{mnras}
\bibliography{biblio}

\bsp	% typesetting comment
\label{lastpage}
\end{document}